\documentclass[a4paper]{jpconf}
\usepackage{graphicx}
\begin{document}

\title{The QGP dynamics in relativistic
heavy-ion collisions}

\author{E. L. Bratkovskaya$^{1,2}$, W. Cassing$^3$, V. P. Konchakovski$^{3,4}$,
O.~Linnyk$^3$, V. Ozvenchuk$^2$, V.D. Toneev$^{5,2}$ and V. Voronyuk$^{5,2,4}$}
\address{$^1$ Institute for Theoretical Physics, University of Frankfurt, Frankfurt, Germany}
\address{$^2$ Frankfurt Institute for Advanced Study, Frankfurt, Germany}
\address{$^3$ Institute for Theoretical Physics, University of Giessen, Giessen, Germany}
\address{$^4$ Bogolyubov Institute for Theoretical Physics, Kiev, Ukraine}
\address{$^5$ Joint Institute for Nuclear Research, Dubna, Russia}

\begin{abstract}
The dynamics of partons and hadrons in relativistic nucleus-nucleus
collisions is analyzed within the novel Parton-Hadron-String Dynamics
(PHSD) transport approach, which is based on a dynamical
quasiparticle model for the partonic phase (DQPM) including a dynamical
hadronization scheme.  The PHSD model reproduces a large variety of
observables from SPS to LHC energies, e.g. the quark-number scaling
of elliptic flow, transverse mass and rapidity spectra of charged
hadrons, dilepton spectra, open and hidden charm production,
collective flow coefficients etc., which are associated with the
observation of a sQGP.  The 'highlights' of the latest results on collective flow are
presented and open questions/perspectives are discussed.
\end{abstract}

\section{Introduction}
The dynamics of the early universe in terms of the 'Big Bang' may
be studied experimentally by ultrarelativistic nucleus-nucleus
collisions at Relativistic-Heavy-Ion-Collider (RHIC) or Large-Hadron-Collider (LHC)
energies in terms of 'tiny bangs' in the
laboratory.  The
Power Spectrum extracted from the Cosmic Microwave Background
Radiation  has some analogy to the Fourier components of particles in the azimuthal
angular distribution \cite{MMSS08}. The discovery of large azimuthal anisotropic flow at
RHIC has provided conclusive
evidence for the creation of dense partonic matter in
ultra-relativistic nucleus-nucleus collisions. With sufficiently
strong parton interactions, the medium in the collision zone can be
expected to achieve local equili\-brium and exhibit approximately
hydrodynamic flow~\cite{Ol92,HK02,Sh09}. The momentum anisotropy is
generated due to pressure gradients of the initial ``almond-shaped''
collision zone produced in noncentral collisions~\cite{Ol92,HK02}.
The azimuthal pressure gradient extinguishes itself soon after the
start of the hydrodynamic evolution, so the final flow is only
weakly sensitive to later stages of the fireball evolution. The
pressure gradients have to be large enough to translate an early
asymmetry in density of the initial state to a final-state
momentum-space anisotropy. In these collisions a new state of
strongly interacting matter is created, being characterized by a
very low shear viscosity $\eta$ to entropy density $s$ ratio,
$\eta/s$, close to a nearly perfect fluid~\cite{Sh05,GMcL05,PC05}.
Lattice QCD (lQCD) calculations~\cite{Cheng08,aori10} indicate that
a crossover region between hadron and quark-gluon matter should have
been reached in these experiments.

An experimental manifestation of this collective flow is the
anisotropic emission of charged particles in the plane transverse to
the beam direction. This anisotropy is described by the different
flow parameters defined as the proper Fourier coefficients $v_n$ of
the particle distributions in azimuthal angle $\psi$ with respect to
the reaction plane angle $\Psi_{RP}$. At the highest RHIC collision
energy of $\sqrt{s_{NN}} =$ 200~GeV, differential elliptic flow
measurements $v_2(p_T)$ have been reported for a broad range of
centralities or number of participants $N_{part}$. For $N_{part}$
estimates, the geometric fluctuations associated with the positions
of the nucleons in the collision zone serve as the underlying origin
of the initial eccentricity fluctuations. These data are found to be
in accord with model calculations that an essentially locally
equilibrated quark gluon plasma (QGP) has little or no
viscosity~\cite{PHEN07,RR07,XGS08}. Collective flow continues
to play a central role in characterizing the transport properties of
the strongly interacting matter produced in heavy-ion collisions at
RHIC and even LHC and shed some light on the scale of initial state fluctuations.

The Beam-Energy-Scan (BES) program proposed at RHIC~\cite{ST11}
covers the energy interval from $\sqrt{s_{NN}}=$ 200~GeV, where
partonic degrees of freedom play a decisive role, down to the AGS
energy of $\sqrt{s_{NN}}\approx$ 5~GeV, where most experimental data
may be described successfully in terms of hadronic
degrees-of-freedom, only. Lowering the RHIC collision energy and
studying the energy dependence of anisotropic flow allows to search
for the possible onset of the transition to a phase with partonic
degrees-of-freedom at an early stage of the collision as well as
possibly to identify the location of the critical end-point that
terminates the cross-over transition at small quark-chemical
potential to a first order phase transition at higher quark-chemical
potential~\cite{Lac07,Agg07}.

This contribution aims to summarize excitation functions for
different harmonics of the charged particle anisotropy in the
azimuthal angle at midrapidity in a wide transient energy range,
{\it i.e.} from the AGS to the top RHIC energy. The first attempts
to explain the preliminary STAR data with respect to the observed
increase of the elliptic flow $v_2$ with the collision energy have
failed since the traditional available models did not allow to
clarify the role of the partonic phase~\cite{NKKNM10}. In this
contribution  we investigate the energy behavior of different flow
coefficients, their scaling properties and differential
distributions (cf.  Ref.~\cite{v2short,v2long}). Our analysis of the
STAR/PHENIX RHIC data -- based on recent results of the BES program
-- will be performed within the Parton-Hadron-String Dynamics (PHSD)
transport model~\cite{PHSD} that includes explicit partonic
degrees-of-freedom as well as a dynamical hadronization scheme for
the transition from partonic to hadronic degrees-of-freedom and vice
versa. For more detailed descriptions of PHSD and its ingredients we
refer the reader to Refs. \cite{BCKL11,Cassing06,Cassing07,Cas09}.

\section{Results for collective flows}
We directly continue with the results from PHSD in comparison with
other approaches and the available experimental data.

\subsection{Elliptic flow}
The largest component, known as elliptic flow $v_2$, is one of the
early observations at RHIC~\cite{Ac01}.
 The elliptic flow coefficient is a widely used quantity characterizing
the azimuthal anisotropy of emitted particles,
 \begin{equation} \label{eqv2}
 v_2 = <cos(2\psi-2\Psi)>=<\frac{p^2_x - p^2_y}{p^2_x + p^2_y}>~,
\end{equation}
where $\Psi_{RP}$ is the azimuth of the reaction plane, $p_x$ and
$p_y$ are the $x$ and $y$ component of the particle momenta and the
brackets denote averaging over particles and events. This
coefficient can be considered as a function of centrality,
pseudorapidity $\eta$ and/or transverse momentum $p_T$. We note that
the reaction plane in PHSD is given by the $(x-z)$ plane with the
$z$-axis in beam direction. The reaction plane is defined as a plane
containing the beam axes and the impact parameter vector.

\begin{figure}[t]
\includegraphics[width=8cm,height=6cm]{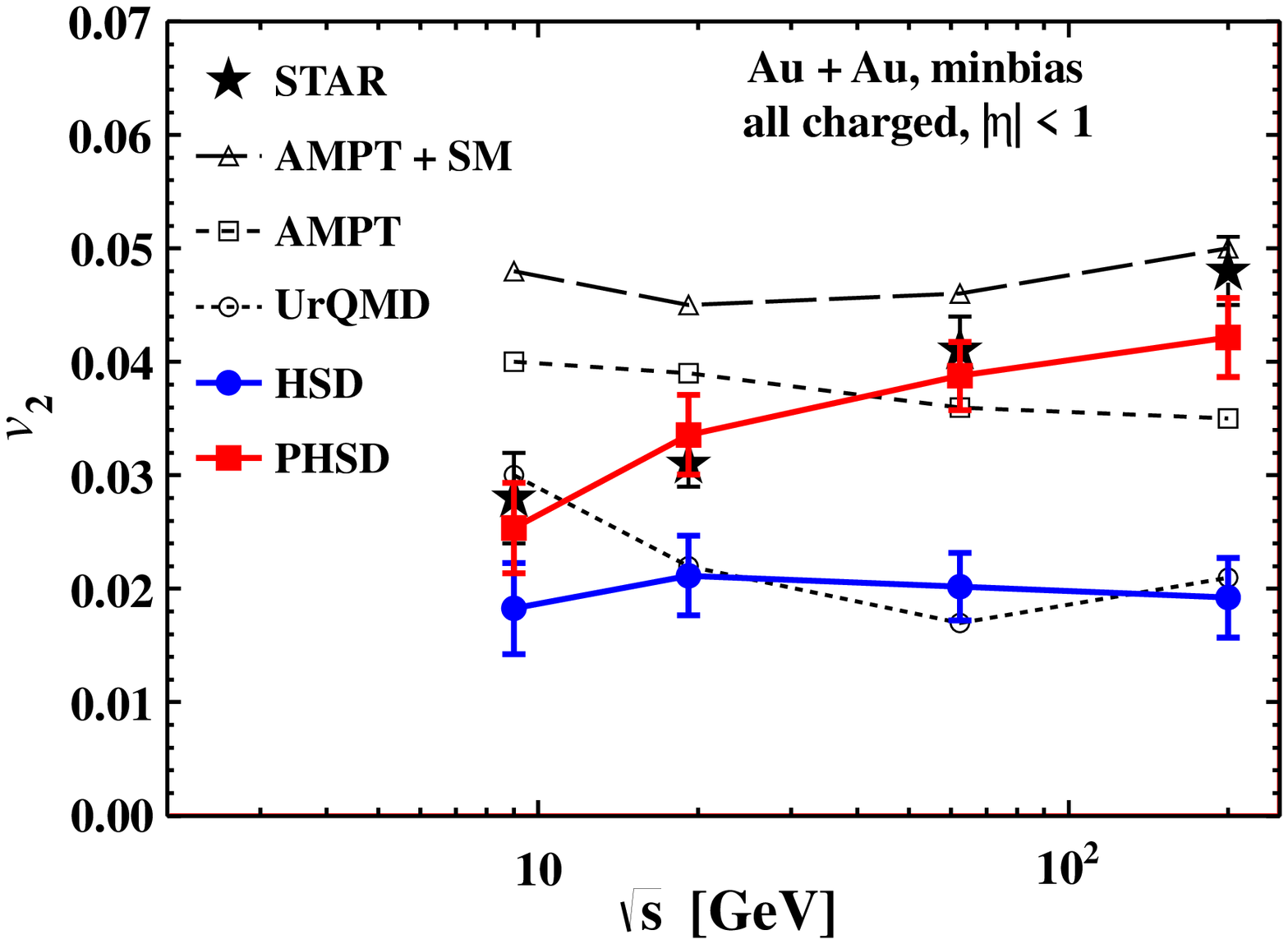}
\includegraphics[width=8cm,height=5.5cm]{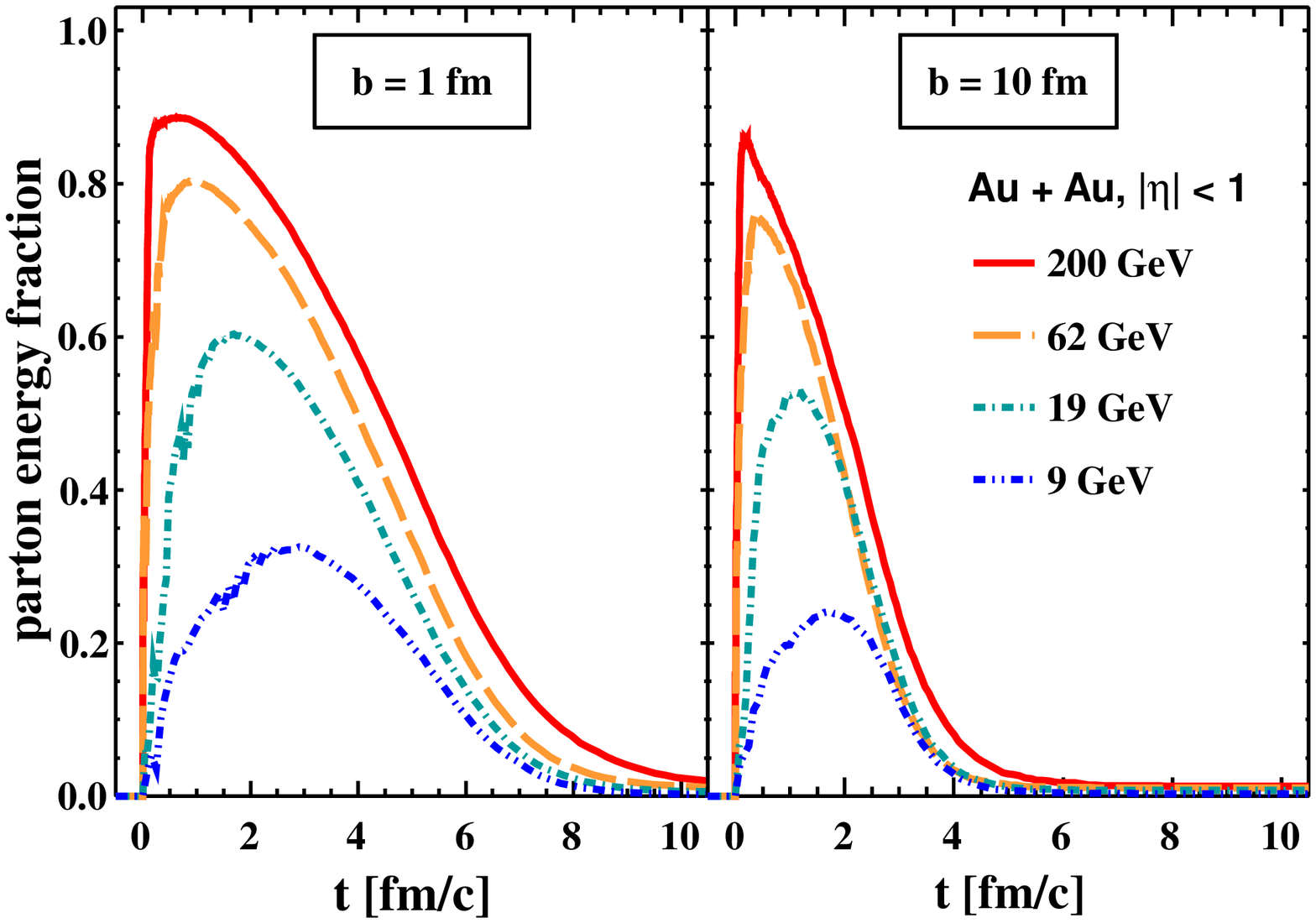}
 \caption{(l.h.s.) The average elliptic flow $v_2$ of
charged particles at
  midrapidity for minimum bias collisions at $\sqrt{s_{NN}}=$ 9.2,
  19.6, 62.4 and 200~GeV (stars) is taken from the data
  compilation of Ref.~\cite{NKKNM10}). The corresponding results from
  different models are compared to the data and explained in more
  detail in the text.  (r.h.s.) Evolution of the parton fraction of the
total energy density
  at midrapidity (from PHSD) for different collision energies at impact parameters
  $b=$1 fm and 10 fm.}
\label{s}
\end{figure}

We recall that at high bombarding energies the longitudinal size of
the Lorentz contracted nuclei becomes negligible compared to its
transverse size. The forward shadowing effect then becomes negligible and the
elliptic flow fully develops in-plane, leading to a positive value
of the average flow $v_2$ since no shadowing from spectators takes
place. In Fig.~\ref{s} (l.h.s.) the experimental $v_2$ data compilation for
the transient energy range is compared to the results from
various models: PHSD, HSD as well as from UrQMD  amd AMPT
as included in Ref.~\cite{NKKNM10}. The centrality selection is the same for the
data and the various models.

The HSD~\cite{BCS03,Ehehalt,HSD} and UrQMD (Ultra relativistic
Quantum Molecular Dynamics) \cite{PR90,UrQMD} are the hadron-string
models and, thus, essentially provide information on
the contribution from the hadronic phase \cite{BRAT04}.
As seen in Fig.~\ref{s}, being in agreement with data at the lowest energy $\sqrt{s_{NN}}=$
9.2~GeV, the HSD and UrQMD model results then either remain approximately
constant or decrease slightly with increasing $\sqrt{s_{NN}}$ and
do not reproduce the rise of $v_2$ with the collision energy
as seen experimentally.

The AMPT (A Multi Phase Transport model)~\cite{AMPT,AMPT2} uses
initial conditions of a perturbative QCD (pQCD) inspired model which
produces multiple minijet partons according to the number of binary
initial nucleon-nucleon collisions.
The string melting (SM) version of the AMPT
model (labeled in Fig.~\ref{s} as AMPT-SM) is based on the idea
of 'melting' of hadrons or strings above the critical energy density
of $\varepsilon\sim$ 1~GeV/fm$^3$ to massless partons. The
subsequent scattering of the quarks are based on a parton cascade
with (adjustable) effective cross sections which are significantly
larger than those from pQCD~\cite{AMPT,AMPT2}.  Once the partonic
interactions terminate, the partons hadronize through the mechanism
of parton coalescence.

We find from Fig.~\ref{s} that the interactions between the minijet
partons in the AMPT model indeed increase the elliptic flow
significantly as compared to the hadronic models UrQMD and HSD. An
additional inclusion of interactions between partons in the AMPT-SM
model gives rise to another 20\% of $v_2$ bringing it into agreement
(for AMPT-SM) with the data at the maximal collision energy. So,
both versions of the AMPT model indicate the importance of partonic
contributions to the observed elliptic flow $v_2$ but do not
reproduce its growth with $\sqrt{s_{NN}}$.
The authors address this
result to the partonic-equation-of state (EoS) employed which
corresponds to a massless and noninteracting relativistic gas of
particles. This EoS deviates severely from the results of lattice
QCD calculations for temperatures below 2-3 $T_c$. Accordingly, the
degrees-of-freedom are propagated without self-energies and a parton
spectral function.

The PHSD approach incorporates the latter medium effects in line
with a lQCD equation-of-state  and also includes a dynamical
hadronization scheme based on covariant transition rates. As has
been demonstrated in Refs. \cite{v2short,v2long} and explicitly
shown in Fig.~\ref{s} (l.h.s.), the elliptic flow $v_2$ from PHSD (red line)
agrees  with the data from the STAR  collaboration and clearly shows
an increase with bombarding energy.

An explanation for the increase in $v_2$ with collision energy is
provided in Fig.~\ref{s} (r.h.s.) where the partonic
fraction of the energy density is shown with respect to the total energy
where the energy densities are calculated at mid-rapidity. As
discussed above the main contribution to the elliptic flow is coming
from an initial partonic stage at high $\sqrt{s}$. The fusion of
partons to hadrons or, inversely, the melting of hadrons to partonic
quasiparticles occurs when the local energy density is about
$\varepsilon\approx$ 0.5~GeV/fm$^3$. As follows from
Fig.~\ref{s}, the parton fraction of the total energy goes down
substantially with decreasing bombarding energy while the duration
of the partonic phase is roughly the same. The maximal fraction
reached is the same in central and peripheral collisions but the
parton evolution time is shorter in peripheral collisions. One
should recall again the important role of the repulsive mean-field
for partons in the PHSD model  that leads to
an increase of the flow $v_2$ with respect to HSD predictions (cf.
also Ref.~\cite{CB08}). We point out in addition that the increase
of $v_2$ in PHSD relative to HSD is also partly due to the higher
interaction rates in the partonic medium because of a lower ratio of
$\eta/s$ for partonic degrees-of-freedom at energy densities above
the critical energy density than for hadronic media below the
critical energy density~\cite{Mattiello,Bass,Vitalii12}. The relative increase
in $v_3$ and $v_4$ in PHSD essentially is due to the higher partonic
interaction rate and thus to a lower ratio $\eta/s$ in the partonic
medium which is mandatory to convert initial spacial anisotropies to
final anisotropies in momentum space~\cite{Pet4}.

\subsection{Higher-order flow harmonics}

Depending on the location of the participant nucleons in the nucleus
at the time of the collision, the actual shape of the overlap area
may vary: the orientation and eccentricity of the ellipse defined by
the participants fluctuates from event to event.  Note, however,
that by averaging over many events an almond shape is regained for
the same impact parameter.
\begin{figure}[t]
\includegraphics[width=8.2cm]{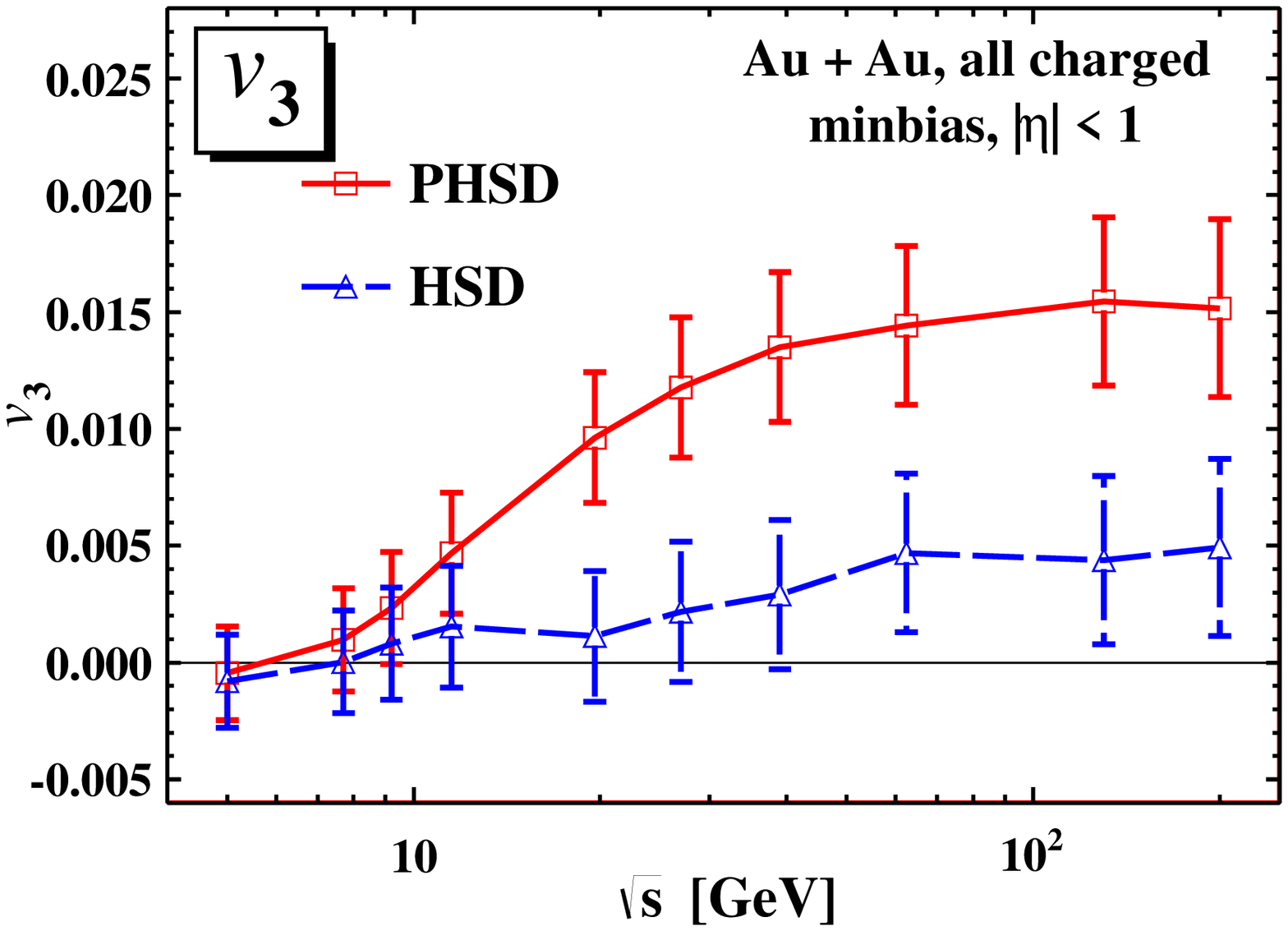}\includegraphics[width=8.2cm]{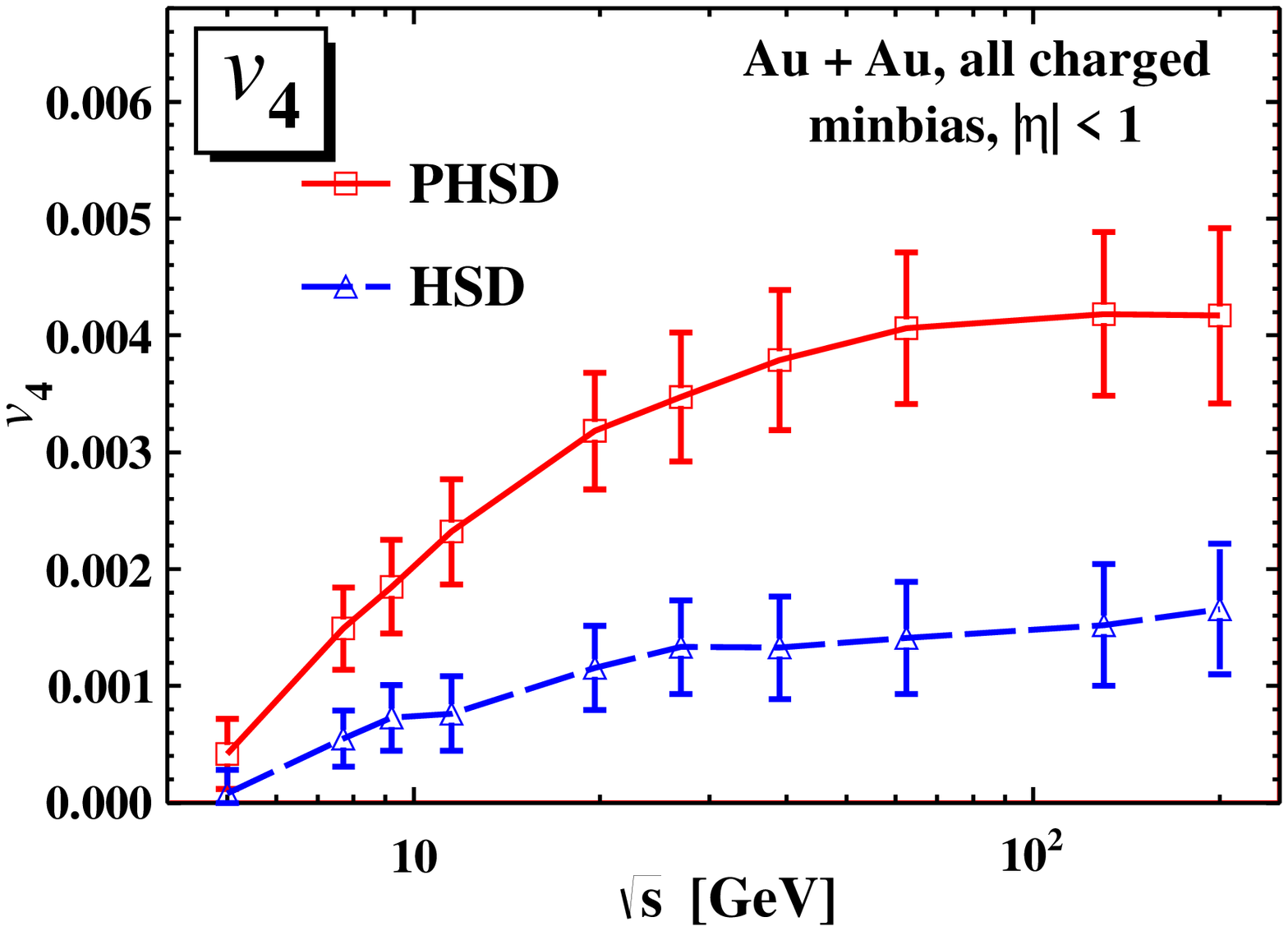}
\caption{Average anisotropic flows $v_3$ (l.h.s.) and $v_4$ (r.h.s.)
of charged
  particles at mid-pseudorapidity for minimum bias Au + Au collisions
  calculated within the PHSD (solid lines, red) and HSD (dashed lines, blue)
  models.}
\label{vns34}
\end{figure}

Recent studies suggest that fluctuations in the initial state
geometry can generate higher-order flow
components~\cite{MMSS08,PHEN07,Pet123,AR10}. The azimuthal momentum
distribution of the emitted particles is commonly expressed in the
form of a Fourier series as
\begin{equation}
E\frac{d^3N}{d^3p}= \frac{d^2N}{2\pi
p_Tdp_Tdy}\left(1+\sum^\infty_{n=1} 2v_n(p_T) \cos
(n(\psi-\Psi_n))\right),\ \ \ \label{eqvn} \end{equation}
where $v_n$ is the magnitude of the $n$-th order harmonic term
relative to the angle of the initial-state spatial plane of symmetry
$\Psi_n$. The anisotropy in the azimuthal angle $\psi$ is usually
characterized by the even order Fourier coefficients with the
reaction plane $\Psi_n=\Psi_{RP}$: $v_n =\langle \exp(\, \imath \,
n(\psi-\Psi_{RP}))\rangle\ ( n = 2, 4, ...)$, since for a smooth
angular profile the odd harmonics vanish. For the odd components,
e.g. $v_3$, one should take into account event-by-event fluctuations
with respect to the participant plane $\Psi_n=\Psi_{PP}$. We
calculate the $v_3$ coefficients with respect to $\Psi_3$ as:
$v_3\{\Psi_3\} = \langle \cos(3[\psi-\Psi_3])\rangle/Res(\Psi_3)$.
The event plane angle $\Psi_3$ and its resolution $Res(\Psi_3)$ are
calculated as described in Ref.~\cite{{AdPH11}} via the
two-sub-events method~\cite{PV98,corrV2}.

In Fig.~\ref{vns34} we display the PHSD and HSD results for the
anisotropic flows $v_3$ and $v_4$ of charged particles at
mid-pseudorapidity for Au+Au collisions as a function of
$\sqrt{s_{NN}}$. The pure hadronic model HSD gives $v_3\approx$ 0
for all energies. Accordingly, the results from PHSD (dashed red
line) are systematically larger than from HSD (dashed blue line).
Unfortunately, our statistics are not good enough to allow for more
precise conclusions. The hexadecupole flow $v_4$ stays almost
constant in the energy range $\sqrt{s_{NN}}\ge$ 10~GeV; at the same
time the PHSD gives noticeably higher values than HSD which we
attribute to the higher interaction rate in the partonic phase.

\begin{figure}[t]
\includegraphics[width=8.1cm]{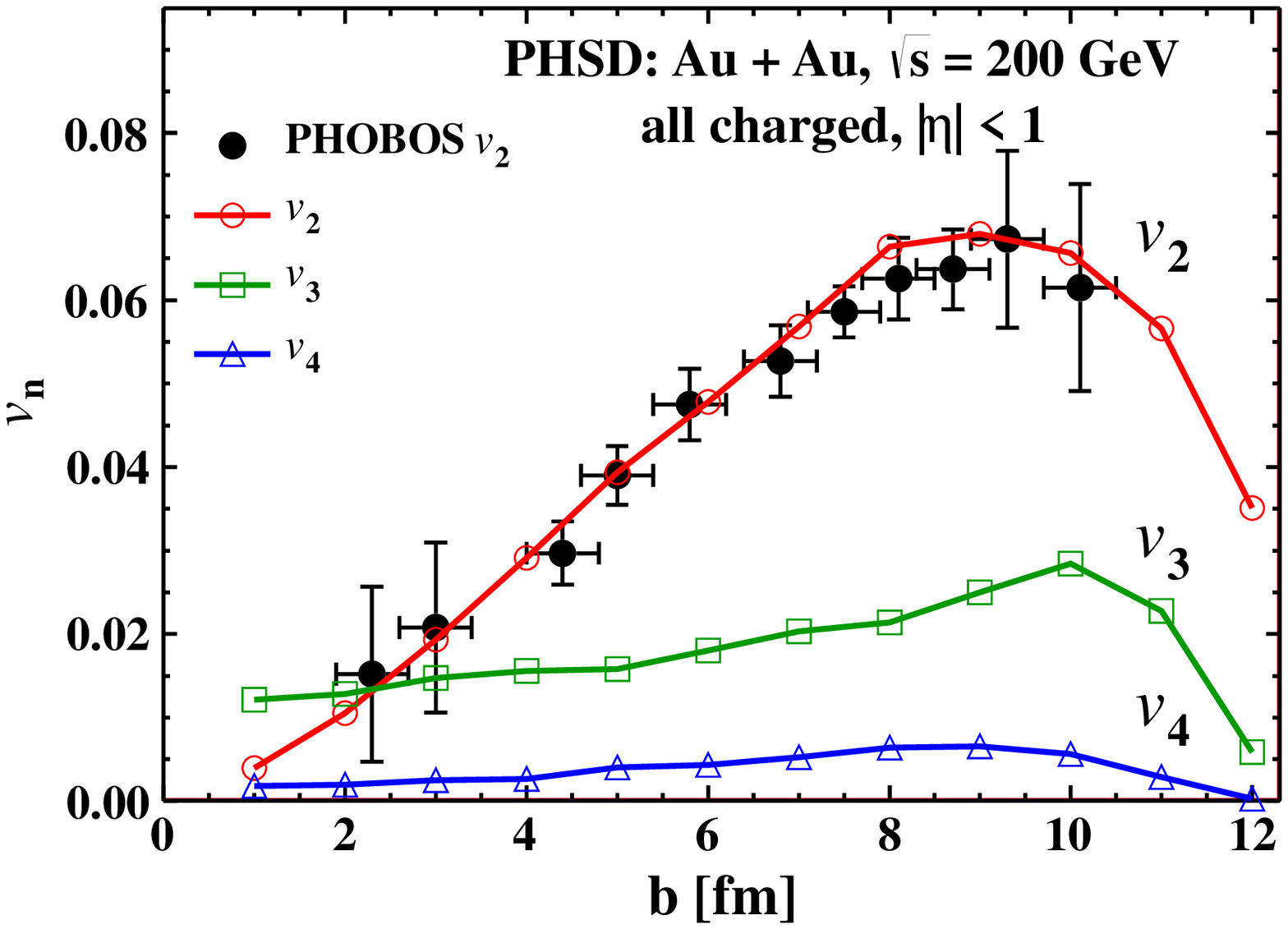}\includegraphics[width=8cm]{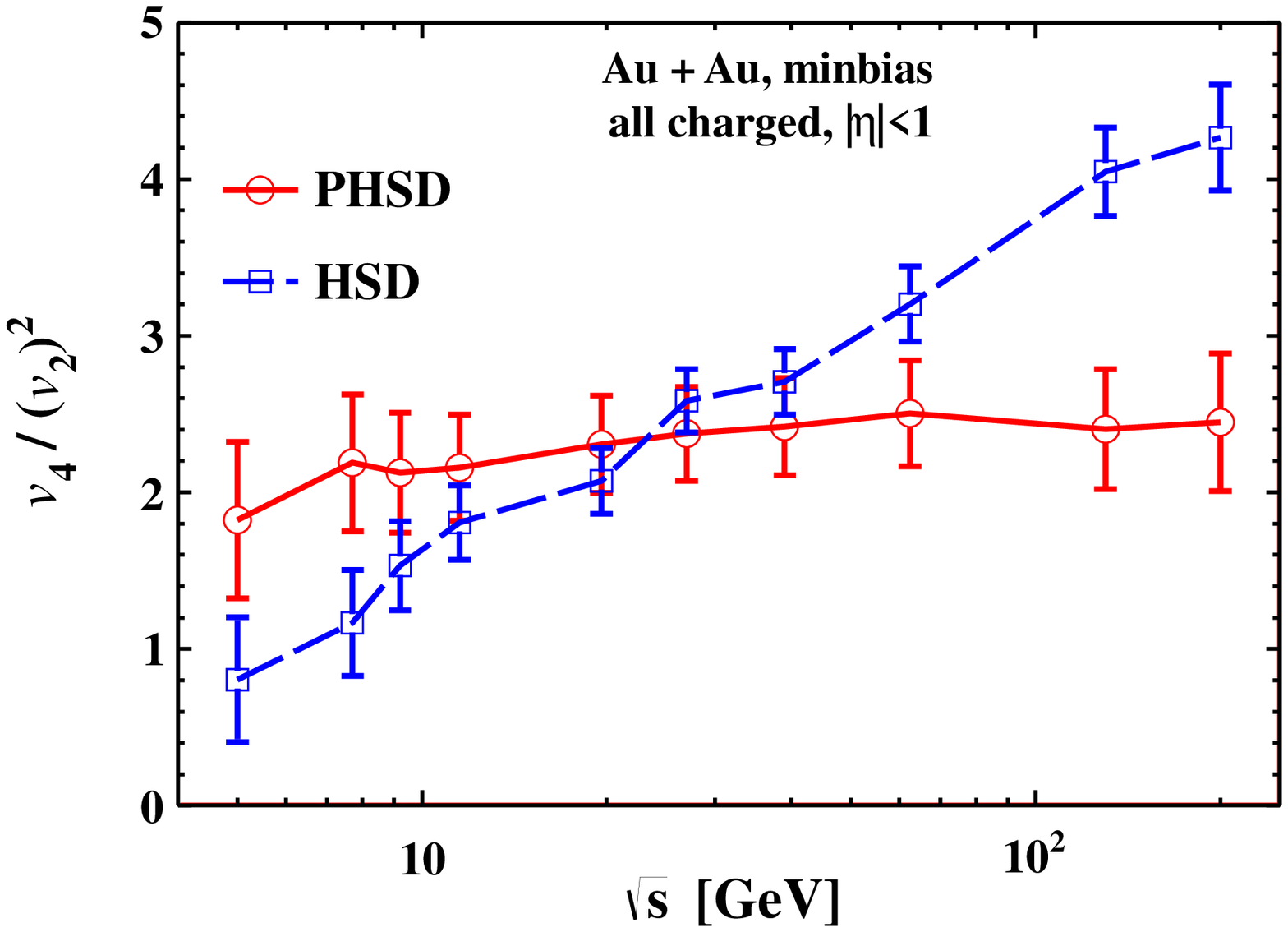}
\caption{(l.h.s.) Impact parameter dependence of anisotropic flows
of charged
  particles at mid-pseudorapidity for minimum bias collisions of Au+Au
  at $\sqrt{s_{NN}}=$ 200~GeV. Experimental points are from
  Ref.~\cite{PHO05}. (r.h.s.) Beam energy dependence of the ratio
$v_4/(v_2)^2$ for Au+Au
  collisions. The solid and dashed curves are calculated within the
  PHSD and HSD models, respectively.} \label{vnb}
\end{figure}

Alongside with the integrated flow coefficients $v_n$ the PHSD model
reasonably describes their distribution over centrality or impact
parameter $b$. A specific comparison at $\sqrt{s_{NN}}=$ 200~GeV is
shown in Fig.~\ref{vnb} for $v_2, v_3$ and $v_4$. While $v_2$
increases strongly with $b$ up to peripheral collisions, $v_{3}$ and
$v_{4}$ are only weakly sensitive to the impact parameter. The
triangular flow is always somewhat higher than the hexadecupole flow
in the whole range of impact parameters $b$.

\subsection{Ratios of different harmonics}
Different harmonics can be related to each other. In particular,
hydrodynamics predicts that $v_4 \propto (v_2)^2$~\cite{Ko03}. The
simplest prediction that $v_4 = 0.5 (v_2)^2$ is given for a boosted
thermal freeze-out distribution of an ideal fluid, Ref.~\cite{BO06}.
In this work it was noted also that $v_4$ is largely generated by an
intrinsic elliptic flow (at least at high $p_T$) rather than the
fourth order moment of the fluid flow. This is a motivation for
studying the ratio $v_4/(v_2)^2$ rather than $v_4$ alone. As is seen
in Fig. 4 (r.h.s.), indeed the ratio calculated within the PHSD
model is practically constant in the whole range of $\sqrt {s_{NN}}$
considered but significantly deviates from the ideal fluid estimate
of 0.5. In contrast, neglecting dynamical quark-gluon
degrees-of-freedom in the HSD model, we obtain a monotonous growth
of this ratio.

\begin{figure}[t]
\includegraphics[width=8.cm]{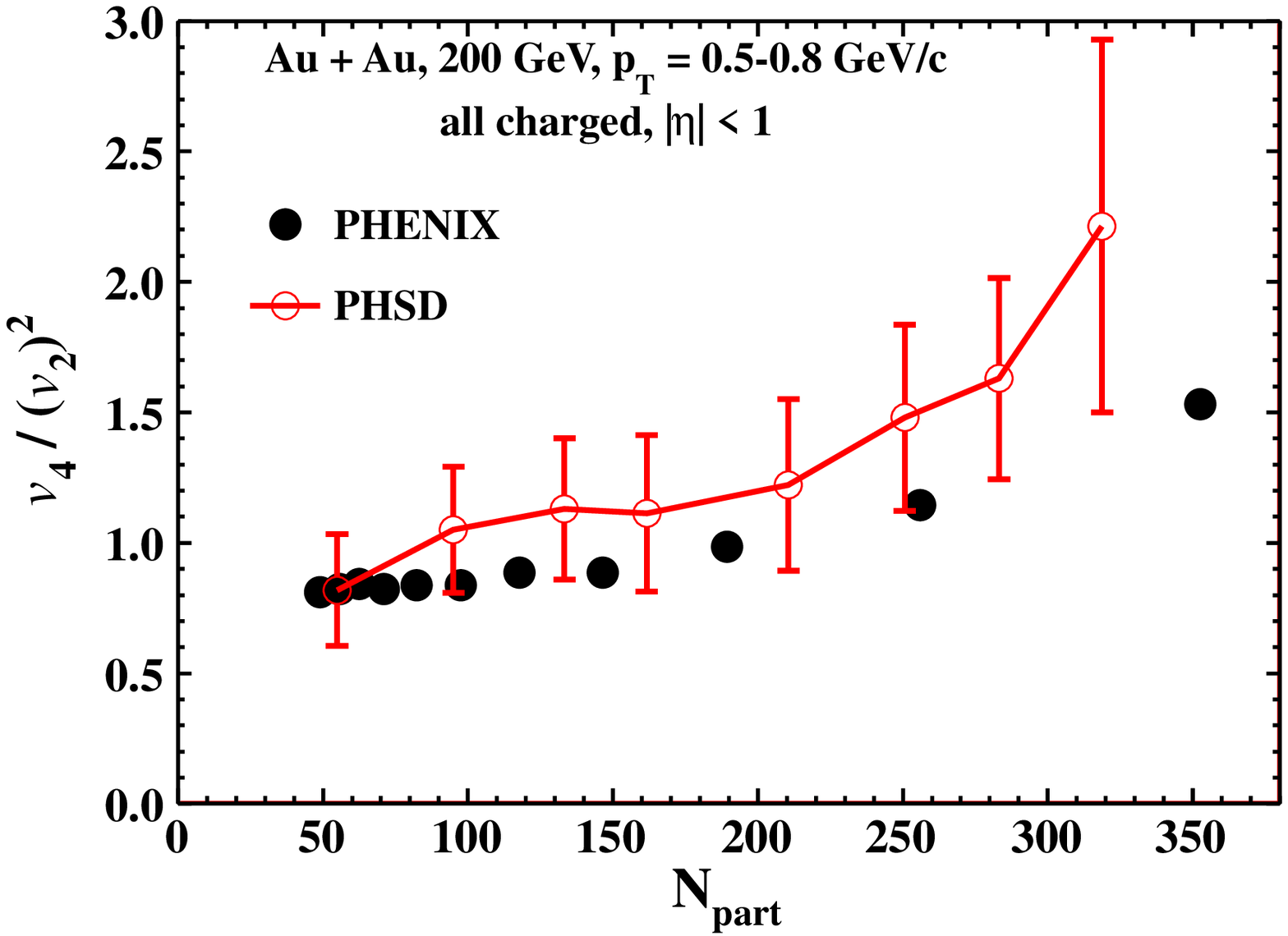}\includegraphics[width=8.cm]{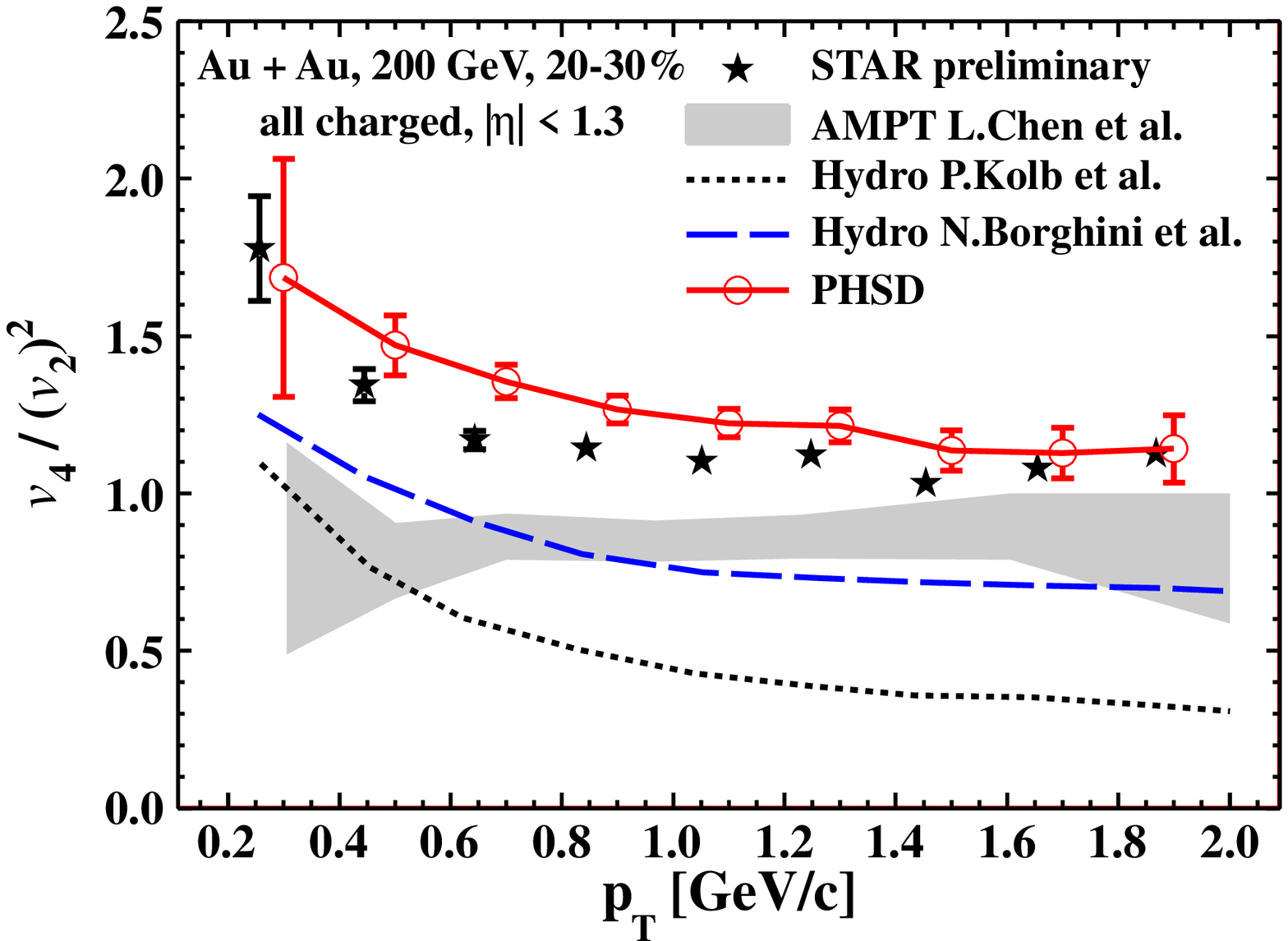}
\caption{(l.h.s.)
 Participant number dependence of the
$v_4/(v_2)^2$ ratio of
  charged particles for Au+Au ($\sqrt{s_{NN}}=$ 200~GeV) collisions.
  The experimental data points for 0.5$<p_T<$0.8~GeV/c are from
  Ref.~\cite{PHENIX-v2-s}. (r.h.s.) Transverse momentum dependence of the ratio
$v_4/(v_2)^2$ of
  charged particles for Au+Au (at $\sqrt{s_{NN}}=$ 200~GeV)
  collisions. The dashed and dot-dashed lines are calculated within
  the hydrodynamic approaches from Refs.~\cite{Ko03} and~\cite{BO06},
  respectively. The shaded region corresponds to the results from the
  AMPT model~\cite{CKL04}. The experimental data points are from the
  STAR Collaboration~\cite{Bai07}.}
\label{v422Np}
\end{figure}

The dependence of the $v_4/(v_2)^2$ ratio versus the number of
participants $N_{part}$ is shown in Fig.~\ref{v422Np} for charged
particles produced in Au + Au collisions at $\sqrt{s_{NN}}=$
200~GeV. The PHSD results are roughly in agreement with the
experimental data points from Ref.~\cite{Bai07} but overshoot them
for $N_{part}\sim$ 250.

As pointed out before, the ratio $v_4/(v_2)^2$ is sensitive to the
microscopic dynamics. In this respect we show the transverse
momentum dependence of the ratio $v_4/(v_2)^2$ in Fig.~\ref{v422Np}
for charged particles produced in Au+Au collisions at
$\sqrt{s_{NN}}=$ 200~GeV (20-30\% centrality). The PHSD results are
quite close to the experimental data points from Ref.~\cite{Bai07},
however, overestimate the measurements by up to 20\%. The
hydrodynamic results -- plotted in the same figure -- significantly
underestimate the experimental data and noticeably depend on
viscosity. The partonic AMPT model~\cite{CKL04} discussed above also
predicts a slightly lower ratio than the measured one, however,
being in agreement with both hydrodynamic models for $p_T\lsim$
0.8~GeV/c. Our interpretation of Fig.~\ref{v422Np} (r.h.s.) is as
follows: the data are not compatible with ideal hydrodynamics and a
finite shear viscosity is mandatory (in viscous hydrodynamics) to
come closer the experimental observations. The kinetic approaches
AMPT and PHSD perform better but either overestimate (in AMPT) or
slightly underestimate the scattering rate of soft particles (in
PHSD). An explicit study of the centrality dependence of these
ratios should provide further valuable information.

\section{Conclusions}
In summary, relativistic collisions of Au+Au from $\sqrt{s_{NN}}=$ 5
to 200~GeV have been studied within the PHSD approach which includes
the dynamics of explicit partonic degrees-of-freedom as well as
dynamical local transition rates from partons to hadrons and also
the final hadronic scatterings. Whereas earlier studies have been
carried out for longitudinal rapidity distributions of various
hadrons, their transverse mass spectra and the elliptic flow $v_2$
as compared to available data at SPS and RHIC
energies~\cite{PHSD,BCKL11}, here we have focussed on the PHSD
results for the collective flow coefficients $ v_2, v_3$ and
$v_4$ in comparison to experimental data in the large energy
range from the RHIC Beam-Energy-Scan (BES) program as well as
different theoretical approaches ranging from hadronic transport
models to ideal and viscous hydrodynamics. We mention explicitly
that the PHSD model from Ref.~\cite{BCKL11} has been used for all
calculations performed in this study and no tuning (or change) of
model parameters has been performed.

We have found that the anisotropic flows -- elliptic $v_2$,
triangular $v_3$, hexadecapole $v_4$ -- are reasonably described
within the PHSD model in the whole transient energy range naturally
connecting the hadronic processes at lower energies with
ultrarelativistic collisions where the quark-gluon degrees of
freedom become dominant. The smooth growth of the elliptic flow
$v_2$ with the collision energy demonstrates the increasing
importance of partonic degrees of freedom. Other signatures of the transverse collective
flow, the higher-order harmonics of the transverse anisotropy $v_3$
and $v_4$ change only weakly from $\sqrt{s_{NN}}\sim$ 7~GeV to the
top RHIC energy of $\sqrt{s_{NN}}=$ 200~GeV, roughly in agreement
with experiment. As shown in this study, this success is related to
a consistent treatment of the interacting partonic phase in PHSD
whose fraction increases with the collision energy.

The analysis of correlations between particles emitted in
ultrarelativistic heavy-ion collisions at large relative rapidity
has revealed an azimuthal structure that can be interpreted as
solely due to collective flow~\cite{
XK11,SJG11,TY10,LGO10}. This
interesting new phenomenon, denoted as triangular flow, results from
initial state fluctuations and a subsequent hydrodynamic-like
evolution. Unlike the usual directed flow, this phenomenon has no
correlation with the reaction plane and should depend weakly on
rapidity. Event-by-event hydrodynamics~\cite{GGH11} has been a
natural framework for studying this triangular collective flow but
it has been of interest also to investigate these correlations in
terms of the PHSD model. We have found the third harmonics to
increase steadily in PHSD with bombarding energy. The coefficient
$v_3$ is compatible with zero for $\sqrt{s_{NN}} >$ 20~GeV in case
of the hadronic transport model HSD which does not develop
`ridge-like' correlations. In this energy range PHSD gives a
positive $v_3$ due to dominant partonic interactions.

Different harmonics can be related to each other and in particular,
hydrodynamics predicts that $v_4 \propto (v_2)^2$~\cite{Ko03}. In
this work it was noted also that $v_4$ is largely generated by an
intrinsic elliptic flow (at least at high $p_T$) rather than the
fourth order moment of the fluid flow. Indeed, the ratio
$v_4/(v_2)^2$ calculated within the PHSD model is approximately
constant in the whole considered range of $\sqrt {s_{NN}}$ but
significantly deviates from the ideal fluid estimate of 0.5. In
contrast, neglecting dynamical quark-gluon degrees-of-freedom in the
HSD model, we obtain a monotonous growth of this ratio.

The transverse momentum dependence of the ratio $v_4/(v_2)^2$ at the
top RHIC energy has given further interesting information ({\it cf.}
Fig. 4) by comparing the various model results to the data from
STAR which are interpreted as follows: the STAR data are not
compatible with ideal hydrodynamics and a finite shear viscosity is
mandatory (in viscous hydrodynamics) to come closer the experimental
 ratio observed. The kinetic approaches AMPT and PHSD perform better but
either overestimate (in AMPT) or slightly underestimate the
scattering rate of soft particles (in PHSD).

We recall that our present PHSD calculations employ 'naturally'
Glauber type initial state fluctuactions which appear compatible
with experimental observations up to top RHIC energies. On the
other hand one might expect that at LHC energies an initial 'glasma' phase
might play a sizeable role and that the gluon-field fluctuations -
of lower scale - could show up in the Fourier decomposition of the
azimuthal angular distribution. It will be interesting to compare
the coefficients $v_n$ for high multiplicity $pp$, $p+Pb$ and
$Pb+Pb$ reactions as a function of transverse momentum $p_T$ (and centrality).

The authors acknowledge financial support through the ``HIC for
FAIR" framework of the ``LOEWE" program.


\section*{References}

\begin{thebibliography}{99}

\bibitem{MMSS08} Mishra A P {\it et al.} 2008 {\it Phys. Rev.} C {\bf 77} 064902

\bibitem{Ol92}
Ollitrault J Y 1992 {\it Phys. Rev.} D {\bf 46} 229

\bibitem{HK02}
Heinz U and Kolb P 2002 {\it Nucl. Phys.} A  {\bf 702} 269

\bibitem{Sh09}
Shuryak E V 2009 {\it Prog. Part. Nucl. Phys.} {\bf 62} 48

\bibitem{Sh05}
Shuryak E V  2005 {\it Nucl. Phys.} A {\bf 750} 64

\bibitem{GMcL05}
Gyulassy M and McLerran L 2005 {\it Nucl. Phys.} A {\bf 750} 30

\bibitem{PC05}
Peshier A and Cassing W 2005 {\it Phys. Rev. Lett.} {\bf 94} 172301

\bibitem{Cheng08}
Cheng M {\it et al.} 2008 {\it Phys. Rev.} D {\bf 77} 014511

\bibitem{aori10}
Aoki Y {\it et al.} 2009 {\it JHEP} {\bf 0906} 088

\bibitem{PHEN07}
Adare A {\it et al.} 2007 {\it Phys. Rev.
Lett.} {\bf 98} 172301


\bibitem{RR07}
Romatschke P and Romatschke U 2007 {\it Phys. Rev. Lett.} {\bf 99} 172301

\bibitem{XGS08}
Xu Z, Greiner C and St\"ocker H 2008 {\it  Phys. Rev. Lett.} {\bf
101} 082302









\bibitem{ST11}
Abelev B I {\it et al.} 2010 {\it Phys. Rev.} C
{\bf 81} 024911

\bibitem{Lac07}
Lacey R A  {\it et al.} 2007 {\it Phys. Rev. Lett.} {\bf 98} 092301

\bibitem{Agg07}
Aggarwal M M  {\it et al.} 2010 {\it arXiv:1007.2613}

\bibitem{NKKNM10}
Nasim M, Kumar L,  Netrakanti P K and Mohanty B 2010 {\it  Phys.
Rev.} C {\bf 82} 054908

\bibitem{v2short}
Konchakovski V P {\it et al.} 2012 {\it Phys. Rev.} C {\bf 85} 011902

\bibitem{v2long}
Konchakovski V P {\it et al.} 2012 {\it Phys. Rev.} C {\bf 85}
044922

\bibitem{PHSD}
Cassing W and Bratkovskaya E L 2009 {\it Nucl. Phys.} A {\bf 831} 215

\bibitem{BCKL11}
Bratkovskaya E L, Cassing W, Konchakovski V P and Linnyk O 2011 {\it
Nucl. Phys.} A {\bf 856} 162

\bibitem{Cassing06}
Cassing W 2007 {\it Nucl. Phys.} A {\bf 791} 365

\bibitem{Cassing07}
Cassing W 2007 {\it Nucl. Phys.} A {\bf 795} 70

\bibitem{Cas09}
Cassing W 2009 {\it Eur. J. Phys.} {\bf168} 3

\bibitem{Ac01}
Ackermann K H {\it et al.} 2001 {\it Phys. Rev. Lett.} {\bf 86} 402

\bibitem{PR90} Cassing W {\it et al.} 1990 {\it Phys. Rep.} {\bf 188} 363

\bibitem{UrQMD}
Bass S A {\it et al.} 1998 {\it Prog. Part. Nucl. Phys.} {\bf 41}
255

\bibitem{BCS03}
Bratkovskaya E L, Cassing W  and St\"ocker H 2003 {\it Phys. Rev.} C
{\bf 67} 054905

\bibitem{Ehehalt}
Ehehalt W and Cassing W 1996 {\it Nucl. Phys.} A {\bf 602} 449

\bibitem{HSD}
Cassing W and Bratkovskaya E L 1999 {\it Phys. Rep.} {\bf 308} 65





\bibitem{BRAT04}
Bratkovskaya E L {\it et al.} 2004 {\it Phys. Rev.} C {\bf 69} 054907

\bibitem{AMPT}
Lin Z W and Ko C M 2002 {\it Phys. Rev.} C {\bf 65} 034904

\bibitem{AMPT2} Lin Z W, Ko C M, Li, B A Zhang B and Pal S 2005 {\it Phys. Rev.} C {\bf
72} 064901



\bibitem{CB08}
Cassing W and Bratkovskaya E L 2008 {\it Phys. Rev.} C {\bf 78} 034919


\bibitem{Mattiello}
Mattiello S and Cassing W 2010 {\it Eur. Phys. J.} C {\bf 70} 243


\bibitem{Bass}
Demir N and Bass S A 2009 {\it Phys. Rev. Lett.} {\bf 102} 172302

\bibitem{Vitalii12} Ozvenchuk V {\it et al.}, {\it arXiv:1212.5393 [hep-ph]}

\bibitem{Pet4}
Petersen H, Coleman-Smith C,  Bass S A and  Wolpert R 2011 {\it J.
Phys.} G {\bf 38} 045102

\bibitem{Pet123} Petersen H and Bleicher M 2010 {\it Phys. Rev.} C
{\bf 81} 044906

\bibitem{AR10}
Alver B and Roland G 2010 {\it  Phys. Rev.} C {\bf 81} 054905

\bibitem{AdPH11}
Adare A {\it et al.} 2011 {\it Phys. Rev. Lett.} {\bf 107} 252301

\bibitem{PV98}
Poskanzer A M and Voloshin S A 1998 {\it  Phys.\ Rev.} C {\bf 58}
1671

\bibitem{corrV2}
Bilandzic A, Snellings R  and Voloshin S A 2011 {\it Phys.\ Rev.} C
{\bf 83} 044913

\bibitem{PHO05}
Back B B {\it et al.} 2005 {\it Phys. Rev.} C {\bf 72} 051901


\bibitem{Ko03}
Kolb P F 2003 {\it Phys. Rev.} C {\bf 68} 031902(R)

\bibitem{BO06}
Borghini N and Ollitrault J Y 2006 {\it Phys. Lett.} B {\bf 642} 227

\bibitem{PHENIX-v2-s}
Gong X Y {\it et al.} 2011 {\it J. Phys.} G {\bf 38} 124146

\bibitem{Bai07}
Bai Y {\it et al.} 2007 {\it J. Phys.} G {\bf 34} S903


\bibitem{CKL04}
Chen L W, Ko C M and Lin Z W 2004 {\it Phys. Rev.} C {\bf 69} 031901

\bibitem{XK11}
Xu J and Ko C M 2011 {\it Phys. Rev.} C {\bf 84} 014903

\bibitem{SJG11}
Schenke B, Jeon S and Gale C 2011 {\it Phys. Rev. Lett.} {\bf 106}
042301

\bibitem{TY10}
Teaney D and Yan L 2011 {\it Phys. Rev.} C {\bf 83} 064904

\bibitem{LGO10}
Luzum M, Gombeaud C and Ollitrault J Y 2010 {\it Phys. Rev.} C {\bf
81} 054910

\bibitem{GGH11}
Gardim F G, Grassi F, Hama Y, Luzum M and Ollitrault J Y 2011 {\it
Phys. Rev.} C {\bf 83} 064901

\end{thebibliography}
\end{document}